\documentclass[twocolumn,noshowpacs,noshowkeys,nofootinbib,aps]{revtex4}
\usepackage{pstricks,graphicx,epsfig,color,amssymb,amsmath,amscd}

\newcommand{\bb}{\bibitem}
\newcommand{\be}{\begin{eqnarray}}
\newcommand{\ee}{\end{eqnarray}}

\begin{document}

\title{On the Brane Cosmology of KK Gravitinos}

\author{Federico R. Urban}

\affiliation{INFN Sezione di Ferrara, via Saragat 1, 44100 Ferrara, Italy\\
Dipartimento di Fisica, Universit\`a di Ferrara, via Saragat 1, 44100 Ferrara, Italy\\
ITEP, Bolshaya Cheremushkinskaya 25, 117218, Moscow, Russia}

\begin{abstract}
In this report I consider the cosmology of KK gravitinos in models with extra dimensions, in particular in connection with the known non--standard high energy regime of expansion which is associated with them. The main result is that the production of such KK modes, once one considers BBN constraints on the allowed entropy released in their decays, is not compatible with non--standard expansion after inflation: there is no five--dimensional Planck mass for which the produced KK gravitinos are safe with respect to BBN. This conclusion holds for both flat and warped models in which only gravity propagates in the full spacetime. This report is based on the work~\cite{our} in collaboration with my colleague Cosimo Bambi.
\end{abstract}

\maketitle

\section{Introduction}\label{intro}

The problem of overproduction of gravitinos, the supersymmetric partner of the graviton, is a long--standing one in cosmology~\cite{problem}. The gravitino interacts very weakly with ordinary matter, its coupling being gravitationally suppressed, and this makes it a long living particle which is never in thermal equilibrium after inflation. 

An unstable gravitino lighter than about 10 TeV decays after the Big Bang Nucleosynthesis (BBN) and the entropy injected into the plasma can cause photodissociation of the light elements, altering their abundances~\cite{abundance}. Hence the requirement of a successful BBN severely constrains the produced amount of gravitinos.

If gravitinos are much heavier, their decay products are not dangerous for primordial nuclei since they harmlessly decay before BBN; however, if $\cal R$--parity is a good symmetry their decay will produce a non--thermal abundance of light SUSY particles, either the lightest stable one (LSP), or some other particle which will later decay into it. The present day energy density stored in LSP as dark matter is constrained by cosmological observations~\cite{wmap}. A similar scenario holds for a light gravitino, lighter than about 100 GeV, which, if it is the LSP as it is likely is the case, must (at least) not overclose the universe. These considerations lead to an upper limit on the temperature at which thermal equilibrium had been established (usually referred to as the reheating temperature $T_R$), this limit being around $10^3\div10^6$ TeV, depending on the model~\cite{abundance,wmap}.

Of course this picture is drastically modified in supersymmetric extra dimensional models~\cite{xsusyA, xsusyB, xsusyC}, in which case one or more KK towers of gravitinos have to be taken into account: it is expected that these extra states will more seriously constrain the allowed maximum temperature reached in the early universe. Furthermore, in braneworlds, where all of the Minimal Supersymmetric Standard Model (MSSM) particles are forced to live on a brane, while (super)gravity is effectively five dimensional, the Friedman equation contains extra terms~\cite{b-cosmology}, which modify the standard cosmological expansion and consequently the picture of gravitino production. Here I report the most important results obtained in~\cite{our} in this context, especially concerning the cosmology of such KK states during non--standard expansion regime.

\section{The gravitino problem}\label{prob}

Gravitinos are produced in several different ways, thermally and non--thermally. Thermal production~\cite{thermal} involves either inelastic scattering processes of thermalised particles, or decays of supersymmetric particles. For the reason explained later, this last mechanism is uninteresting in braneworld cosmology. Non--thermal mechanisms~\cite{non-thermal} include perturbative and non--perturbative production by means of inflaton decay, or some other scalar fields (moduli, dilaton, radion), which is strongly model dependent, and is not treated here. The last option is gravitational particle production, which, for the brane cosmology scenario, is discussed in~\cite{us}.

The zero mode gravitino abundance is usually expressed in terms of the gravitino number density to the entropy density ratio as
\be\label{Ydef}
Y^0_{3/2}(T) = \frac{n^0_{3/2}(T)}{s(T)}\, .
\ee
Here $s(T) = (2\pi^2/45) g_{*S} T^3$ and $g_{*S}$ is the number of ``entropic'' degrees of freedom, and $T$ is the temperature of the system.

The Boltzmann equation for the process under examination leads to the abundance of the thermally produced particles
\be\label{boltzdef}
\frac{d}{dT} Y^0_{3/2} = - \frac{s \langle \sigma v \rangle\, Y_{rad}^2}{H T}\, ,
\ee
where $Y_{rad}$ is the equilibrium number density to entropy density ratio for relativistic particles, and $\langle \sigma v \rangle$ parametrises the thermally averaged cross section for the process under scrutiny. $H$ is the Hubble parameter which in standard cosmology is $H = (\rho/3M_4^2)^{1/2}$, where $\rho$ is the energy density of the universe and $M_4=2.4\cdot10^{15}{\rm TeV}$ is the reduced four dimensional Planck mass. In the radiation dominated epoch of the early universe $\rho \propto T^4$.

Integrating this equation one finds the well known~\cite{abundance} expression for the abundance at the BBN (given that the zero mode is not too heavy and thus survives at least till $T\simeq1$ MeV)
\be\label{zeroYstd}
Y_{3/2}^0 = 1.9 \cdot 10^{-19}\, \left( 1 + \frac{\tilde m^2}{3 {m_{3/2}^0}^2} \right)\, \left( \frac{T_R}{\rm TeV} \right)\, ,
\ee
where $\tilde m$ is the gluino mass, assumed to be below the reheating temperature.

\subsection{The 4D gravitino problem}\label{stdsol}

Since the primordial gravitino abundance (\ref{zeroYstd}) is proportional to the reheating temperature, cosmological constraints on $Y_{3/2}^0$ translate into upper bounds on $T_R$ and hence on the inflationary model. If the gravitino is stable, its energy density today must not overclose the universe. In particular, it must not exceed the dark matter energy density. This puts a bound on $T_R$ only for $m_{3/2} \gtrsim 1$ keV. On the other hand, if the gravitino is unstable its decay products can alter BBN predictions and/or the CMBR spectrum. The resulting constraint depends on several unknown parameters, such as gravitino lifetime and branching ratio. The details are given in~\cite{our,abundance}.

For instance, in the typical case where the gravitino is not the LSP and $m_{3/2} \sim 100 \; {\rm GeV} \div 1 \; {\rm TeV}$, the reheating temperature must be
\be\label{example-T}
T_R \lesssim 10^5 - 10^8 \; {\rm GeV}
\ee
and several inflationary models have to be rejected or strongly fine--tuned, their typical energy scales being around the Planck or GUT scales.

\subsection{The 5D solution}\label{branesol}

So far only the standard cosmological expansion law has been considered. However, if we happened to live on a four dimensional Friedman--Robertson--Walker hypersurface (the brane), embedded in an extra dimensional spacetime, the early universe would admit an epoch of non--standard expansion~\cite{b-cosmology}. Several such models have been built in the last few years, the (representative) ones which are dealt with here being the ADD~\cite{add} and RS~\cite{rs} models, which involve flat and warped extra dimensions respectively. These models show a peculiar feature when their cosmology is investigated. Although the details are model--dependent, the typical and general resulting effect is that the Friedmann equation shows a different high--energy behaviour of this kind~\cite{b-cosmology}:
\be\label{modFried}
H^2 = \frac{\rho}{3M_4^2} \left( 1 + \frac{\rho}{2\lambda} \right)\, ,
\ee
where $\lambda$ is the tension of the brane, which is related to the five dimensional Planck mass as $\lambda=6M_5^6/M_4^2$. This equation says that at high energy densities the expansion of the universe was much faster than at later times, and went as $T^4$ instead of $T^2$, together with the unknown parameter $M_5$: the smaller $M_5$ the faster the expansion.

At this point it is convenient to define a ``transition'' temperature $T_*$ from standard cosmology to brane one, which can be extracted from $\rho=2\lambda$~\cite{osamu}
\be\label{trT}
T_*^2 = 
\left(\frac{360}{\pi^2 \, g_*}\right)^{1/2}\, \frac{M_5^3}{M_4}\, ,
\ee
where $g_* = g_*(T)$ counts the relativistic degrees of freedom at a given temperature $T$. If the dominant component of the universe is not radiation then this ``temperature'' approximately means the fourth root of the energy density, and parametrises the epoch at which the transition occurs.

The new expansion law needs to be taken into account when the amount of gravitino produced in the early universe is calculated, that is, (\ref{modFried}) has to be plugged into (\ref{boltzdef}). Under the assumptions that $T_R\gg T_*$ and $T_*\gg T$, and that the extra dimension does not change the coupling of the gravitino zero mode to the matter residing on the brane, instead of (\ref{zeroYstd}) the abundance at the BBN is approximately given by~\cite{osamu}
\be\label{zeroYbrane}
Y_{3/2}^0 = 3.5 \cdot 10^{-19}\, \left( 1 + \frac{\tilde m^2}{3 {m_{3/2}^0}^2} \right)\, \left( \frac{T_*}{\rm TeV} \right)\, .
\ee

The main point here is that the constraints on $T_R$ need now to be imposed on $\sim2T_*$, and thus on the unknown five dimensional mass scale, while allowing for any reheating temperature after inflation, as high as we wish. This would of course be very good news, if it were not for the fact that, as we will see in a minute, extension of the analysis to the full spectrum of KK modes leads to a much worse (although somewhat model--dependent) situation.

\section{SUSY and extra dimensions}\label{xsusy}

Supersymmetry and supergravity in the context of extra dimensions has been investigated by several authors, primarily in connection with supersymmetry breaking by means of extra dimensional mechanisms~\cite{xsusyA, xsusyB, xsusyC}. The main reason of interest on these models revolves around superstring theory, for it requires both supersymmetry and extra dimensions, although the path from such low--energy models and the full underlying string theory is far from being crystal clear. The cosmology of these models has not been studied yet, and, while it is expected that the well known main features of brane cosmology still hold, even relevant modifications could arise, primarily due to extra field in the bulk (the gravitino) and model--dependent orbifolding boundary conditions. This possibility is not explored further here, as the following analysis is readily extended to other cosmologies.

Here I will shortly report the important formulas which are needed in the calculations, while I refer to the original paper~\cite{our} for the detailed calculations.

\subsection{Flat bulk}\label{flatB}

In this model the bulk spacetime is flat and contains only gravitons and gravitinos. In considering the non--standard expansion epoch only the model with one extra dimension is analysed, since in this case the modified Friedmann equation (\ref{modFried}) holds, whereas little is known for the general case.

The mass for each state can be expressed as\footnote{Henceforth the $n$--th KK gravitino mass will be just $m_n$.}~\cite{xsusyA}
\be
m_n &=& m_0 + \frac{n}{R}\label{mADD} \, .
\ee

Here $R$ is the size of the extra dimension, while $m_0$ is the zeroth mass, which can be either fixed by the extra dimensional parameters (this is the case if SUGRA is broken thanks to a mechanism which relies on the extra dimensions themselves), or not~\cite{xsusyA}. Since there is no agreement on the way supergravity is broken, the zero mode mass will be taken as a free parameter. That specified, the mass gap between two nearby states is given by
\be\label{gapADD}
\Delta m = \frac{1}{R} = \frac{2\pi M_5^3}{M_4^2} = 
\left( \frac{\pi^4 g_*}{90} \right)^{1/2}\, \frac{T_*^2}{M_4}\, .
\ee

Coming to the coupling constants, the situation is tricky and highly model dependent. In this report I will be using the standard parametrisation for the cross section extracted from (\ref{zeroYstd}), where of course the $n$--th KK gravitino mass $m_n$ has to be taken into account, however see~\cite{our} for more details.

\subsection{Warped bulk}\label{warpB}

The second model to be dealt with is the warped one. Now a five dimensional cosmological constant resides in the bulk, which makes it an $AdS_5$ spacetime. Once again, the Friedmann equation receives a high--energy correction as in (\ref{modFried}).

The mass spectrum is discrete, the KK modes masses being given by the following formula
\be\label{mRS}
m_n = m_0 + k x_n e^{- \pi k R}\, ,
\ee
where $x_n$ is a solution of $J_1(x_n) = 0$ ($J_1$ is the BesselJ function of the first kind), $k$ is the $AdS_5$ curvature
\be
k = \frac{M_5^3}{M_4^2} \left(1 - e^{- 2 \pi k R}\right)\, ,
\ee
and $R$ parametrises the size of the extra dimension. In this case as well, the mass splitting is independent on the way SUSY is broken, as it follows directly from the extra dimensional setup~\cite{rs, rs-tower}. The mass gap reads
\be\label{gapRS}
\Delta m	&=& k e^{ -\pi k R} \left( x_n - x_{n-1} \right) \simeq 3 k e^{ -\pi k R} \nonumber\\
					&=& \left (\frac{\pi^2 g_*}{40} \right)^{1/2} \frac{1 - e^{ -2\pi k R}}{e^{ \pi k R}}\, \frac{T_*^2}{M_4} \nonumber\\
					&\equiv& \left (\frac{\pi^2 g_*}{40} \right)^{1/2} F(k R)\, \frac{T_*^2}{M_4}\, ,
\ee
where $F(k R)$ is defined by the last equality.

The coupling constants will again be taken to be the usual ones, and I will only consider a single KK tower of gravitinos, while in the actual scenario it is likely that two or more towers have to be taken into account. However, since $N=1$ $D=4$ SUGRA already forbids non--standard expansion, it is not necessary to consider other supergravities: they would make the picture even worse, see~\cite{our}.

\section{The 5D gravitino problem revisited}\label{branekk}

Gravitinos are initially produced during a high temperature era. The total abundance for a given KK mode is computed by integrating (\ref{boltzdef}), where the $n$--th mode mass has to be taken into account. The upper limit for the integral is the highest temperature reached in the early universe for which the relativistic plasma was in thermal equilibrium; the lower limit is the temperature at which thermal production stops, which, for each mode, is approximately equal to its mass.

The abundance generated so far remains constant, except for some small jumps in the total entropy density, until it is time for these gravitinos to decay. If the expansion always followed the standard Friedmann law, the number density to entropy density ratio for the $n$--th gravitino mode at the BBN would be
\be\label{nthYstd}
Y_{3/2}^n	= 1.9 \cdot 10^{-19}\, \left( 1 + \frac{\tilde m^2}{3 m_n^2} \right)\, \left( 1 - \frac{m_n}{T_R} \right) \left( \frac{T_R}{\rm TeV} \right)\, .
\ee

At this point both the zero mass and the mass gap are unspecified, hence, the calculation of the total amount of gravitinos could involve either an integral over the relevant range of masses, which is from $n = 0$ to $n = (T_R - m_0) / \Delta m \simeq T_R / \Delta m$, or a summation over them. In the continuum limit, for standard cosmological expansion, the result is (notice that the continuum approximation might not be always valid~\cite{our}),
\be
Y_{3/2}^{\rm tot}	&\simeq& 10^{-19}\, \left( \frac{T_R}{\rm TeV} \right) \{ \frac{T_R}{\Delta m} +\nonumber\\
									&+& \frac{2 \tilde m^2}{3 \Delta m^2} \frac{\Delta m}{m_0} \left( 1 + \frac{m_0}{T_R}\, \ln\frac{m_0}{m_0 + T_R} \right)\; \} \label{totYint1} \nonumber\\
									&\simeq& 10^{-19}\, \left( \frac{T_R}{\rm TeV} \right) \{ \frac{T_R}{\Delta m} + \frac{2 \tilde m^2}{3 \Delta m^2} \frac{\Delta m}{m_0}\; \} \label{totYint2} \, .
\ee

The history of these gravitinos is in principle very complicated as it strongly depend upon their masses and couplings, but it is reasonable to consider the following approximation~\cite{our}. The KK gravitino tower could be split into four ``bands'', keeping in mind that the lightest band may not exist if the zero mode is heavy enough. The first band consists of the modes for which\footnote{The next--to--LSP (NLSP) mentioned here is not the first KK gravitino, but the lightest non--gravitino MSSM particle.} $m_n < m_{NLSP}$: once produced they will remain as non--thermal relics. If there are direct transitions between KK modes only those for which their lifetime is longer than the age of the Universe contribute to the dark matter today, but their abundance is fed by the decays of heavier modes. It will be seen that these light modes are not going to be very relevant, though. The second band is that for which $m_{(N)LSP} < m_n < m_{MAX} \simeq 10^5 \; {\rm TeV}$: these gravitinos end up as out--of--equilibrium LSP relics, whoever the LSP is. In the third band superheavy gravitinos live: they either decay into thermalised particles, and contribute nothing to non--thermal relics abundances today, or decay into lighter KK modes, increasing their abundances and tightening the constraints following from non--thermal gravitinos. The fourth band, which overlaps the second, and possibly also the first one, is the band for which gravitino decays affect BBN: this band may admit less freedom for the parameters of the models, and goes approximately from 100 GeV to 30 TeV.

We wish now to generalise the previous discussion allowing for an epoch of non--standard expansion, which would have taken place after $T_R$ but before BBN. The Friedmann equation is given by (\ref{modFried}). The $n$--th mode abundance is thus given by (again the gluino term in the cross section is neglected):
\be\label{nthYbrane}
Y_{3/2}^n	&\simeq& 10^{-19} \left( \frac{T_R}{\rm TeV} \right) \phantom{x}_2F_1[ \frac{1}{4} , \frac{1}{2} ; \frac{5}{4} ; -\left( \frac{T_R}{T_*} \right)^4] \nonumber\\
					&\simeq& 3 \cdot 10^{-19} \left( \frac{T_*}{\rm TeV} \right)\, ,
\ee
where $F$ is the Gauss Hypergeometric function, and the last step implies $T_R \gg T_*$. This equation basically means that gravitinos are mainly produced around $T_*$, regardless of $T_R$ as long as it is much bigger than $T_*$ itself. This is the result obtained for the zero mode in~\cite{osamu}.

\subsection{Flat extra dimension}\label{ADDbrane}

Since all the gravitinos with masses lighter than $T_*$ are produced in the amount predicted by (\ref{nthYbrane}), while the production of heavier ones is strongly suppressed, the total gravitino abundance will be given by
\be\label{totYbraneADD}
Y_{3/2}^{\rm tot} \simeq 10^{-3} \frac{T_*^2}{M_4 \Delta m} \simeq 10^{-4}\, ,
\ee
where the last equality follows from (\ref{gapADD}) and is valid for a flat fifth extra dimension. This is the first central result: it is straightforward to conclude that KK gravitinos and the non--standard expansion epoch are not compatible with each other, the only possible way out being that all of the KK masses lie outside the range for which gravitinos are constrained by overclosure or BBN, but this seems unrealistic since it would require fine tuning of the zeroth mass together with $T_*$.

This can be seen in another way: the available number of KK states, inversely proportional to $T_*^2$, grows faster than the amount which can be cut away by lowering $T_*$ itself. Thus, once a small $T_*$ is taken, as demanded by the zeroth gravitino bound, many KK states would become available below that temperature, which would require a further step downwards for $T_*$, which in turn implies even more KK states available, and so on. There is no value for $M_5$ for which a safe enough amount of gravitinos is produced. This is entirely due to the relation between $T_*$ and $\Delta m$.

For instance, had the transition temperature been chosen around $10^5$ TeV, as imposed by the zeroth gravitino constraint, the mass gap would have been around $3 \cdot 10^{-5}$ TeV, which means an enormous number (about $10^9$) of KK states available at that temperature. 

\subsection{Warped extra dimension}\label{RSbrane}

In this case one could hope that since the mass gap depends on two unknown parameters there will be some parameters space for which the conclusion of the previous section could be evaded. However, despite this fact, unless the gravitino zeroth mass and the temperature scales on the scene are finely tuned, there is still no way one can get rid of the too many KK gravitinos.

In the warped case the mass gap is given by (\ref{gapRS}), and the overall amount of gravitinos becomes
\be\label{totYbraneRS}
Y_{3/2}^{\rm tot} \simeq 10^{-3} \frac{T_*^2}{M_4 \Delta m} 
\simeq \frac{10^{-3}}{F(k R)} \, .
\ee

One can easily see that this case is not better than the previous one, since the function $1/F(k R)$ is always bigger than approximately 2.5. This means that $Y_{3/2}^{\rm tot} > 3 \cdot 10^{-3}$, which is of course too much. If the zeroth gravitino constraint is imposed one would find $\Delta m / {\rm TeV} \simeq 10^{-5} F(k R) \lesssim 3 \cdot 10^{-6}$: tons of KK states are available in this scenario as well.

\subsection{On KK gravitons}\label{gravitonsSec}

Up to now the focus has been mainly on gravitinos. It is quite natural to wonder whether similar conclusions could be deduced by considering gravitons alone, whose better known properties furnish more reliable grounds for discussing BBN constraints~\cite{gravitons}.

In ADD--like models, KK graviton interactions are $1/M_4^2$ suppressed, so, the corresponding lifetime is basically the same of KK gravitinos of equal mass, as long as we are not dealing with gravitino--goldstino states. However, since the graviton zero mode is massless and gravitons are not supersymmetric particles (and thus they have not a corresponding graviton {\cal R}--parity), only those whose masses lie within $100 \; {\rm GeV} \div 30 \; {\rm TeV}$ are useful. Indeed, heavier gravitons provide essentially no bounds, because their decay can not affect BBN or produce stable and dangerous relic particles. Concerning lighter gravitinos, only fairly weaker constraints can be deduced from BBN and CMBR, because their decay could spoil BBN predictions and/or produce distortions of the CMBR spectrum. Nevertheless, KK gravitons in the mass range $100 \; {\rm GeV} \div 30 \; {\rm TeV}$ should suffice, that is, KK gravitons and non--standard expansion in ADD--like models are not compatible as well.

The situation is completely different in RS--like models, as here KK graviton wavefunctions are peaked on ``our'' brane, so they interact much more strongly. In this case KK gravitons could thermalise, and no relevant bounds would be obtained. Of course if some modes do not reach thermal equilibrium, they would be able to constrain the reheating (or transition) temperature, even though these limits are expected to be much more shallow than what has been obtained here.

So, to conclude with a single statement: KK gravitinos and non--standard expansion cannot be arranged at the same time without upsetting, in particular, BBN, and this conclusion is definitive only when KK gravitinos (in addition to KK gravitons) are considered explicitly.

\section{Conclusion}\label{end}

We have considered the phenomenology of toy models where supergravity is realised and subsequently broken in an extra dimensional setup, and we studied the cosmology of KK gravitino states which arise in that case. The most relevant conclusion is that, unless a considerable fine tuning between masses and parameters of the extra dimensional model is required, it is not possible to allow for an epoch of non--standard expansion and, at the same time, avoid KK gravitino overproduction. This is true for both flat and warped extra dimensional models, as long as there is at least one weakly interacting tower of KK gravitinos.

As far as high (${\cal O}(100)$ GeV or more) temperatures are concerned, these results are relatively general, as they do not rely on $\pm 1/2$ states which are significantly model--dependent. In regard to light KK gravitinos, general predictions are rather difficult to be made, since there is not a complete model of supersymmetry in extra dimensions.

Importantly, similar bounds coming from the KK tower of gravitons are not so tight. First of all, KK gravitons, if weakly interacting, provide constraints only if they decay after BBN, that is, only for a given range of masses, whereas KK gravitinos would produce stable particles (LSP) and must be demanded to not exceed the observed amount of cosmological dark matter. Secondly, in warped models KK gravitons interact strongly and in the early universe they thermalise, while KK gravitinos should not. Thus, much stronger limits can be obtained by investigating KK gravitinos cosmology, especially in warped models.

{\bf Acknowledgments --} I wish to thank all the organisers of the ITEP winter school 2008 for the splendid environment in which the school has taken place, and to everyone at ITEP, and to Misha Vysotsky in particular, for the hospitality during my whole winter stay at ITEP.

\end{document}